\documentclass[preprint,showkeys,aps,prb]{revtex4}
\usepackage[dvips]{graphicx}
\begin{document}

\title{
Random Walk Equivalence to the Compressible Baker Map
and the Kaplan-Yorke Approximation to Its Information Dimension
}

\author{
William Graham Hoover and Carol Griswold Hoover                                    \\
Corresponding Authors' emails : hooverwilliam@yahoo.com \& hoover1carol@yahoo.com  \\
Ruby Valley Research Institute                                                     \\
Highway Contract 60, Box 601, Ruby Valley, Nevada 89833, USA ;                     \\
}

\date{\today}


\vspace{0.1cm}

\begin{abstract}

Simple deterministic model systems, with {\it time-reversible} equations of motion, can generate
{\it irreversible} phase-space flows with attractor-repellor pairs satisfying the Second Law of
Thermodynamics. Maps, and equivalent random walks, can also do this. To illustrate this paradoxical
reversibility situation we study a pair of time-reversible contractive Baker Maps, $N2$ and $N3$. Both generate
dissipative {\it fractal} phase-space structures. The steadily decreasing phase-space volumes exhibited
by iterating these maps correspond to the dissipation associated with entropy production. Like three
smooth reversible dissipative one-body phase-space flows developed in the 1980s and 1990s these
simple two-dimensional maps generate fractal distributions, but in two dimensions rather than three,
simplifying visualization and analyses. The continuity equation, which quantifies phase-volume loss,
motivates study of the fractals' reduced ``information dimensions'', which were approximated by
Kaplan and Yorke in terms of two-dimensional maps' two Lyapunov exponents. The maps studied here
generate fractal (fractional dimensional) distributions in their phase spaces. By mapping uniformly
dense grids of points, fractal dimensions can be determined by ``area-wise'' mappings. Beginning
with a uniform grid area-wise mapping of the $N2$ Baker Map provides an information dimension of 1.78969.
Alternatively, as many as a trillion iterations, starting from an arbitrary point, gives a smaller
``point-wise'' dimensionality, $1.741_5$. Neither of these precisely determined estimates matches
the Kaplan-Yorke conjecture value, 1.7337. In the course of studying these three different approaches
to information dimension we developed random walk equivalents to both mappings, which greatly simplifies
analyses. We found that for the older $ N2$ Baker map the three approaches all disagree with one another!
We later discovered that for the newer $N3$ Baker mapping the three approaches to information dimension,
area-wise, point-wise and Kaplan-Yorke, agree.
\end{abstract}

\maketitle

\section{Numerical Simulations of ManyBody Dynamics}

Statistical mechanics, developed in the 19th and early 20th centuries by Boltzmann in Austria, Gibbs
in the United States, and Maxwell in England, provides a formalism giving macroscopic thermodynamic properties in
terms of microscopic $(q,p)$ phase-space trajectory properties. But the complexity of systems more
complicated than the ideal gas or the harmonic crystal prevented much progress on ``realistic'' manybody
problems in particle or astrophysical dynamics. By the mid-20th century computers played a huge role in
designing weapons for World War II. Their ability to solve complex problems quickly caught the attention
of physicists, mathematicians, engineers, chemists, ... , all of whom were stymied by the complexity of
their nonlinear equations in many variables. After the war computers could be applied to many of the ``hard
problems'' that had accumulated as fruits of the scientific revolution. Computer simulations of manybody
problems were developed at universities and national laboratories worldwide. Straightforward applications
of particle mechanics and statistical mechanics stimulated international collaborations long before email
could make such cooperations routine.

As a result of 1980s and 1990s workshop and conference meetings in Berlin, Budapest, Gmunden, New Hampshire,
Orsay, Warwick, and Zakopane, Bill, with half a dozen colleagues, developed several one-body toy-model small
systems designed to shed light on the simulation of (irreversible) nonequilibrium systems with time-reversible
equations of motion\cite{b1,b2,b3,b4,b5}. Among the research goals of these scientists were the resolutions of two
paradoxes which had puzzled Maxwell and Boltzmann and their followers, Loschmidt's, a consequence of
time-reversible motion equations:
\begin{quote}
 ``How can time-reversible motion equations simulate irreversible processes?''
\end{quote} and Zerm\'elo's, a consequence of the Poincar\'e recurrence of any dynamical state in a
bounded portion of $(q,p)$ phase space:
\begin{quote}
``How can entropy only increase if the initial state will inevitably recur?''
\end{quote}

Applications of two computational innovations combined to provide resolutions of these paradoxes. In the mid-1980s
Shuichi Nos\'e developed a revolutionary variant of Hamiltonian dynamics\cite{b6,b7}. He introduced a control
variable, his ``time-scaling variable'', influencing the kinetic temperature. This modified dynamics, still time-reversible,
enabled the simulation of systems at a specified kinetic temperature rather than constant energy. This work was improved
and simplified by Bill Hoover\cite{b8,b9} as a result of conversations he and Nos\'e had near the Notre Dame Cathedral in 1984. They
had met by chance at a train station in Paris, a few days prior to a CECAM workshop in Orsay. By 1986 Nos\'e-Hoover dynamics was
generalized to the simulation of nonequilibrium steady states. Bill, along with half a dozen colleagues, developed three
toy-model problems illustrating applications of the new mechanics' temperature control to three nonequilibrium systems: the
Galton Board\cite{b2}, the Galton Staircase\cite{b1,b3}, and, a decade later, the Conducting Oscillator\cite{b5}. The three problem
types all exhibited irreversible chaotic solutions (exponentially sensitive to perturbations) despite the deterministic  
time-reversibility of the dynamics. [ 1 ] The Galton Board problem follows the field-driven isokinetic
motion of a hard disk through a fixed lattice of identical hard-disk scatterers. The resulting phase-space
distribution is fractal\cite{b2,b10}, a distribution with a nonintegral topological dimensionality. [ 2 ] The Galton
Staircase problem likewise follows a thermostatted field-driven motion, but of a mass point with momentum $p$in a
sinusoidal potential. The equations of motion for the Galton Staircase are
$$
\dot q = p \ ; \ \dot p = F - \sin (q) - \zeta p \ ; \ \dot \zeta = p^2 - T \ .
$$
[ 3 ] The Conducting Oscillator problem\cite{b5} simulates the motion of a heat-conducting harmonic oscillator thermostatted
with a coordinate-dependent temperature $T(q) = 1 + \epsilon \tanh (q)$.

All three of these Nos\'e-Hoover modifications of Hamiltonian flows can generate fractal distributions and do also
obey the phase-space continuity equation expressing the comoving conservation of probability $fdqdpd\zeta= f\otimes$.
Here $f$ is the probability density and $\otimes$ is an infinitesimal phase volume element:
$$
(\dot f/f) = -(\dot \otimes/\otimes) =
-[(\partial \dot q/\partial q)+(\partial \dot p/\partial p)+(\partial \dot \zeta /\partial \zeta)]
 =  \zeta = (\dot S/k) \ .
$$
Gibbs' and Boltzmann's identification of entropy with $\langle -k\ln f \rangle $ identifies the Nos\'e-Hoover friction
coefficient $\zeta$ with entropy production. This is a useful result in interpreting nonequilibrium
simulations including the instantaneous heat transfer to the external heat baths represented by the
temp0erature-control variable $\zeta = (\dot S/k)$. Here $k$ is Boltzmann's constant.  For convenience we usually
choose it equal to unity.

In these three deterministic time-reversible models thermostatting is implemented by integral feedback forces
imposing a given kinetic temperature $\langle \ p^2 \ \rangle$, with control forces $\{-\zeta p\}$
linear in the moving particle's momentum $p$.  These model systems are sufficiently simple that their
phase-space distributions can be analyzed precisely\cite{b10,b11} to determine the power-law variation of phase-space
bin probabilities P($\delta$) with bin size $\delta$. The resulting box-counting and correlation
dimensionalities of the fractal distributions describe the scaling of the zeroth and second powers of bin
probabilities $\{$ P $\}$. The information dimension is logarithmic. It corresponds to
$\langle \ln$(P)$\rangle / \ln (\delta)$, giving the powerlaw variation of the density of points with respect to the bin size.
Information dimension arises naturally in analyzing thermostatted mechanics and is the focus of our attention here.
Because one-, two-, and three-dimensional objects in a three-dimensional space have probabilities varying as the
first, second, and third powers of the bin size $\delta$ the definition of the information dimension,
$D_I=\langle \ln ({\rm P}) \rangle/\ln (\delta)$, is a natural generalization of dimension from the integers 1, 2,
3 to a continuously variable ``fractal'' value. In the special toy-model cases studied in the 1980s and 1990s most
distributions turned out to have fractional rather than integral dimensionalities, characteristic of nonequilibrium
steady states. Under some conditions one-dimensional dissipative limit cycles resulted\cite{b5}.

\section{Time-Reversible Chaos and the Two-Dimensional Baker Map}
Solutions of Hamilton's or Lagrange's or Newton's or Nos\'e-Hoover's motion equations are all ``time-reversible''.
A transparent example is the solution of the one-dimensional harmonic oscillator with unit mass and force constant;
$$
\dot q = p \ ; \ \dot p = -q \rightarrow \ddot q = - q \ [ \ {\rm Hamiltonian \ Oscillator} \ ] \ ;
$$
$$
\ddot x = -x(t) \ [ \ {\rm Newtonian \ Oscillator} \ ] \ ;
$$
$$
\ddot x = -x(t)-\zeta \dot x \ ; \ \dot \zeta = \dot x^2 - T \ [ \ {\rm Nos\acute{e}-Hoover \ Oscillator} \ ] \ .
$$

Given initial values of the coordinate, $x$ or $q$, at the current and previous times, $x(t)$ and $x(t-dt)$, one can
integrate either forward or backward, extending the coordinates' time series as far into the future or past
as desired. Time reversibility can be confirmed by integrating for one timestep, changing the sign of $dt$
and integrating (backward in time) for one step, and then again changing the time, returning to the initial values
of $x(t),\dot x(t),\zeta(t)$ or $(q(t),p(t))$. Adding a Nos\'e-Hoover thermostatting force $-\zeta p$ the dynamics retains
time-reversibility so long as $\zeta$ changes sign in the reversed motion, behaving like a momentum variable\cite{b9}.

Studies of chaotic flows require at least three dynamical variables. In a bounded region of one-or-two-dimensional
space a deterministic trajectory must either stop or trace out a periodic orbit, and so cannot be chaotic. The
graphics can be simplified by considering projections or cross-sections of three-dimensional flows. A little reflection
shows that cross-sections of flows are equivalent to maps, with deterministic finite jumps from one phase-space point to another
rather than a smooth continuous flow. Let us consider the reversibility of maps. Textbook maps were typically both
dissipative and irreversible in 1987\cite{b1}. At that time Bill had no idea that maps could be time-reversible.
He wrote\cite{b1}:
\begin{quote}
``The mathematical structures of dissipative maps and the hydrodynamic equations are inherently irreversible. The
Nos\'e-Newton equations are different: They are time-reversible.''
\end{quote}

\section{Generating Time-Reversible Baker Maps}
\begin{figure}
\includegraphics[width=2.3 in,angle=+90.]{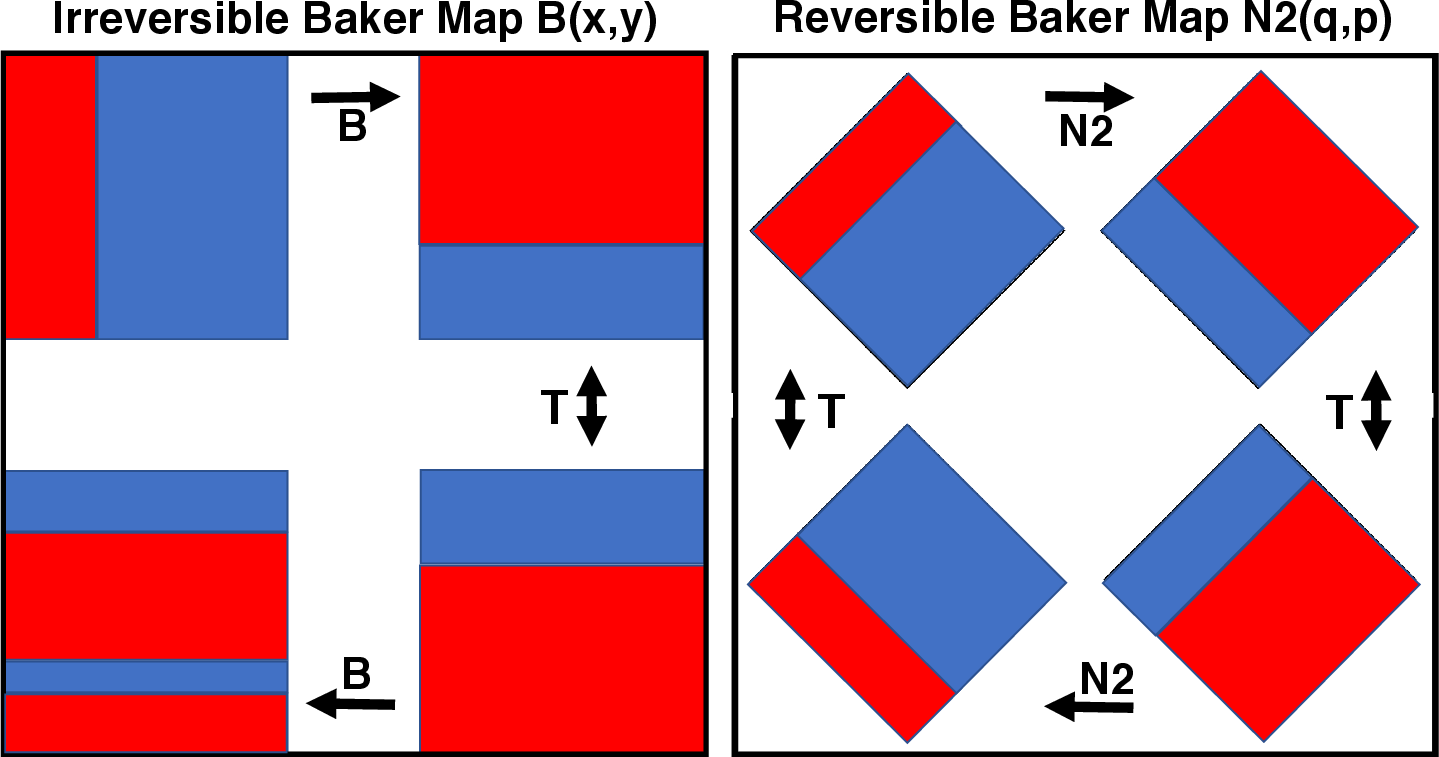}
\caption{
The two-panel $B(x,y)$ (at left) and $N2(q,p)$ (at right) versions of the compressible nonequilibrium Baker
Map. For convenience the mapping is illustrated in the unit square, $0<x,y<1$ at left and in a
$2\times2$ diamond at the right with $-\sqrt{2}< q,p < +\sqrt{2}$. In both cases the mapping $T$, not to be
confused with temperature,
changes the sign of the vertical coordinate, $T(\pm x,\pm y)= (\pm x,\mp y)$ at the left and
$T(\pm q,\pm p)= (\pm q,\mp p)$ at the right. Note that the bottom leftmost configuration differs
from a time-reversed image of the top left image, showing that map $B$ is not time-reversible. The 45
degree rotated mapping $N2$ at the right satisfies time reversibility $N2TN2T(q,p) = I(q,p) = (q,p)$, and
so is a more faithful analog of time-reversible classical mechanics. Here $I$ is the identity mapping.
}
\end{figure}

\begin{figure}
\includegraphics[width=2.2 in,angle=+90.]{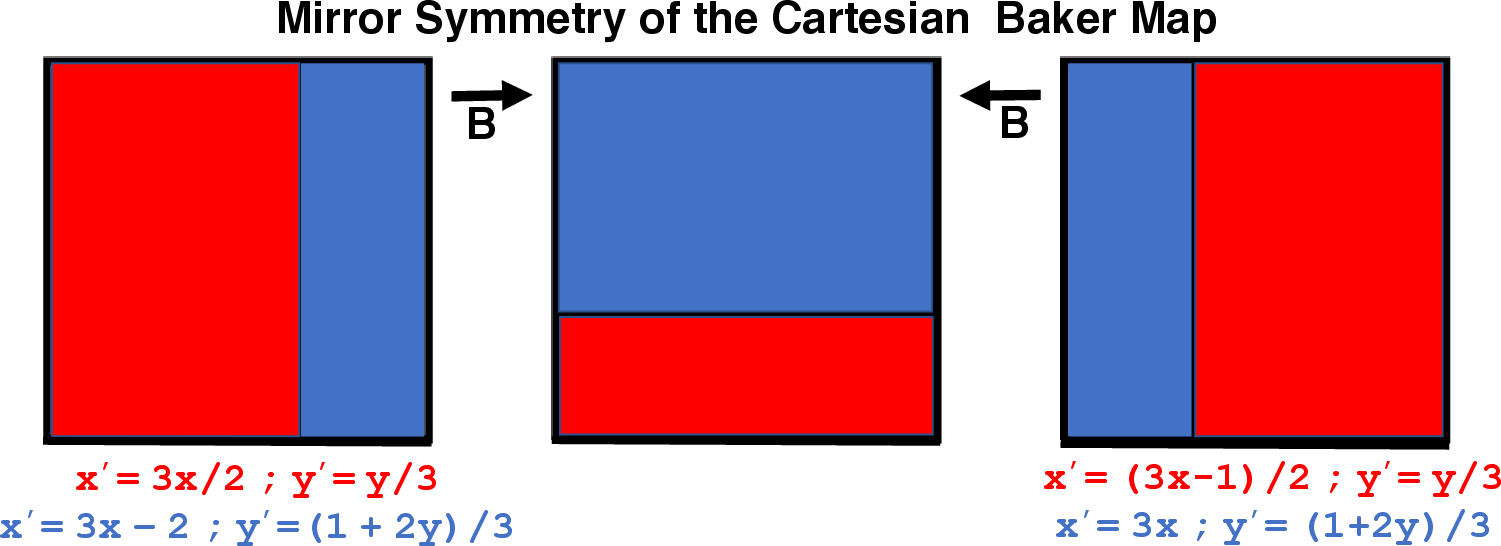}
\caption{
The mirror symmetry of the Baker map implies that the dividing line between the two mappings, contracting
the red region and expanding the blue, can be located at $x = 1/3$ (at the right, above) or $x = 2/3$ (at
the left). The analytic forms of the red and blue mappings are colored accordingly. The red mappings halve
the area while the blue mapping expand, in both cases by a factor of two. The primed coordinates describe
coordinates in the central unit square.
}
\end{figure}

\begin{figure}
\includegraphics[width=2.1 in,angle=-90.]{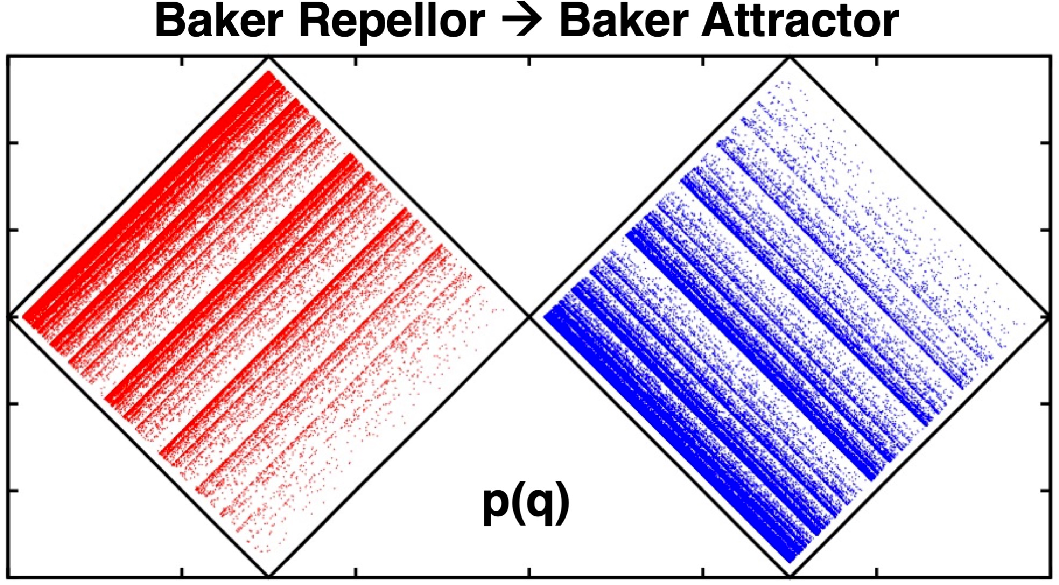}
\caption{
100,000 iterations of the inverse $N2^{-1}$ of the nonequilibrium Baker Map $N2(q,p)$ generate
the fractal repellor (red, at the left). Changing the sign of the vertical ``momentum''
$p$ generates the fractal attractor (blue, at the right) from the repellor. Point-wise analyses
of either fractal with trillions of iterations suggest an information dimension $D_I = 1.741_5$.
The mappings shown here were achieved ``pointwise'', by repeated mappings of a single point.
The limiting extrapolated steady-state information dimensions of the two fractals, based on
large-$n$ meshes of width $(1/3)^n$, are close to 1.741, as is discussed in the text. The
Kaplan-Yorke Lyapunov dimension is significantly smaller, 1.7337 for the two rotated Baker maps.
For a related puzzle see {\bf Figure 3}.
}
\end{figure}

If a {\it time-reversible} map $M(q,p)$ maps a coordinate $q$ and momentum $p$ forward for one step then it must obey
the identity $I=MTMT$, where $T$ changes the sign of the momentum $p$ and $I$ is the identity,
$$
I(q,p) = (q,p) \ ; \ M(q,p) = (q',p') \ ; \ T(q,\pm p) = (q,\mp p) \ .
$$
We choose the left-to-right convention, 123\dots, for the ordering of sequences of mappings. For instance,
with M time-reversible, the sequence of four mappings $MTMT$ corresponds first to stepping forward with $M$,
second to shifting to reverse, third to stepping backward with $M$, and fourth, changing the direction of
motion from reverse to forward, matching the original direction of motion, $MTMT=I$.

Reversibility can be implemented by considering the rotational modification $N2$ of the Baker's Map $B$,
shown at the right in {\bf Figure 1}. This modification clears the way for area changes corresponding to
the production of Boltzmann-Gibbs' entropy. The two-panel Baker map $N2$ (at the right) doubles the size
of an area element $dxdy$ in the red region at upper left and halves that of an element from the larger
blue region. The two mappings (one for red points and one for blue) are linear, with the ``new'' coordinate
or momentum of the form $A + Bq + Cp$. The constants $(A,B,C)$ can be identified relatively easily from the
mappings of the vertices of like-colored regions. See the example equations in {\bf Figure 2}.

The resulting mappings for the two-panel Baker maps can be expressed as follows:
In conventional Cartesian coordinates, with $0 < x,y < 1$ the Cartesian map $B$ for red elements of area
in the top row {\bf Figure 1} is
$$
x < (1/3) \longrightarrow x' = 3x \ ; \ y' = (1+2y)/3 \ [ \ {\rm Red \ Mapping} \ ] \ .
$$
Blue elements likewise follow a linear mapping:
$$
x > (1/3) \longrightarrow x' = (3x-1)/2 \ ; \ y' = y/3 \ [ \  {\rm Blue \ Mapping} \ ] \ .
$$
To check the reversibility of these maps simply apply the combination $BTBT$ to the vertices and check to
see whether or not the original points are recovered. Because the combination mapping $BTB$ produces four
parallel horizontal strips rather than two vertical strips at the lower left of {\bf Figure 1} the
Cartesian Baker Map $B$ (at top left) is {\it not} time-reversible.

By analogy with flows a map $M$ is said to be time-reversible when it can be reversed by a three-step
process: [ 1 ] changing the signs of the momentum-like variables, [ 2 ] propagating all the variables
one (``backward'') iteration, and then changing the signs of the momenta once more, so that the inverse
of the map $M$ is given by $M^{-1} = TMT$. In ordinary Hamiltonian mechanics the $T$ mapping simply maps
$(\pm q,\pm p) \rightarrow (\pm q,\mp p)$.  Bill's conversations with Bill Vance and Joel Keizer during
Vance's graduate work at the University of California's Davis campus led us to a nonequilibrium
{\it rotated} version of the Baker Map $B$ which we call $N2$, for ``Nonequilibrium''with two panels.
This Map's domain is the diamond-shaped region, centered on $(q,p)=(0,0)$ and shown at the right of
{\bf Figure 1} and again in {\bf Figure 3}. Now imagine that the map $N2$ is applied to a representative
input point $(q,p)$. This operation produces the next point $(q^\prime,p^\prime)$.

Our rotated nonequilibrium Map, $N2(q,p)\rightarrow (q^\prime,p^\prime)$ has the following analytic form :
\noindent
For (red) twofold expansion, $q < p - \sqrt{2/9}$ :
$$
q' = (11q/6) - (7p/6) + \sqrt{49/18} \ ; \ p' = (11p/6) - (7q/6) - \sqrt{25/18} \ .
$$
For (blue) twofold contraction, $q > p - \sqrt{2/9}$ :
$$
q' = (11q/12) - (7p/12) + \sqrt{49/72} \ ; \ p' = (11p/12) - (7q/12) - \sqrt{1/72} \ .
$$
{\bf Figure 3} shows the resulting concentration of probability into bands parallel to the
attractor's bottom left and the repellor's upper left edges of their diamond-shaped domains.

Although the algebra is more cumbersome we have chosen to use the rotated $N2(q,p)$ version
of this map, centered on the origin and confined to a diamond-shaped region of
sidelength 2, as shown at the right in the {\bf Figures}. We regard the horizontal $q$ variable
as a coordinate and the vertical $p$ variable as a momentum.  {\bf Figures 1 and 3} illustrate
the time-reversibility of the $(q,p)$ map. This similarity to nonequilibrium molecular
dynamics, along with the square roots generating the $45^o$ rotation, are twin
advantages of this nonequilibrium diamond-shaped map $N2$.  The square roots eliminate
most of the artificial periodic orbits resulting from finite computer precision. Beginning at the
center point of the Cartesian rational-number square map, $(x,y)=(0.5,0.5)$, leads to a periodic
orbit of just 3095 single-precision iterations. Starting instead at the equivalent central point of
the irrational-numbered diamond map, $(q,p)=(0.0,0.0)$, leads to a single-precision periodic orbit of
1,124,068 iterations. With double-precision arithmetic the orbits are much longer.  $10^{12}$ such 
$(x,y)$ iterations from the same initial condition gave no repeated points. Let us next consider an
approximate theoretical approach to analyzing the Baker fractal followed by two computational
approaches. We will find several interesting surprises in so doing.

\section{Kaplan and Yorke's Conjectured Dimension}

It has been argued\cite{b11} that the fractal information dimension is best suited to characterizing
fractal distributions of points because it is uniquely insensitive to changes of variables.
For that reason Kaplan and Yorke's conjectured relation between the Lyapunov spectrum and the
information dimension, $D_{KY} = 1 -(\lambda_1/\lambda_2) \stackrel{?}{=}D_I$ in this case, is of
special interest. Because the Baker Map is linear one might expect that
it would likely follow the conjectured relation. Kaplan and Yorke suggested that a linear interpolation
formula between the number of terms in the last positive sum of exponents, starting with the largest,
$\lambda_1$, and the number of terms in the next sum ( the first negative sum, one greater than the number
of terms in the previous sum),
would be a useful estimate for the information dimension\cite{b12} . In fact they cite many a case,
including theoretical work carried out by L. S. Young, for which their conjectured estimate is exactly
correct.

The blue portion of the compressible Baker Map of $B$ in {\bf Figure 1} represents the (2/3) of the
measure that stretches horizontally by a factor (3/2) while the red portion represents that (1/3) of
the measure that  stretches by a factor of 3 in the same direction, horizontally. As a result the
longtime stretching rate per iteration is
$$
\lambda_1 = (2/3)\ln (3/2) + (1/3)\ln (3) = (1/3)\ln (27/4) = 0.63651 \ .
$$
Likewise (2/3) of the measure shrinks vertically by a factor 3 as does (1/3) by a factor (2/3) so that                                                        
$$
\lambda_2 = (2/3) \ln (1/3) + (1/3) \ln (2/3) = (1/3)\ln(2/27) = -0.86756 \ .
$$
The linear interpolation between the single-term ``positive sum'', 0.63651, and the two-term sum, $0.63651-0.86756=
-0.23105$, gives an interpolated ``number of terms for a sum of zero'', 1 + ($0.63651/0.86756) = 1.73368$.
This dimension, sometimes called the ``Lyapunov dimension'' {\it is} the Kaplan-Yorke dimension $D_{KY}$.

In their 1998 paper\cite{b4}, presented at the 1997 Budapest Meeting on Chaos and Irreversibility\cite{b16},
 Bill and Harald Posch introduced the two-panel nonequilibrium $N2$ Baker map. The model stimulated more work
at the meeting\cite{b17} and subsequently\cite{b18}. In 2005 Kum\^c\'ak wrote a very readable paper\cite{b19} emphasizing the
connection of ``Generalied Baker maps'' to the phase-space contractability (to fractals) providing improved
understanding of the emergence of the Second Law of Thermodynamics for such models. Kumi\^c\'ak characterized his
generalized maps with the variable $w$. The fraction of a mapping occupied by the narrowest strip, is $1/w$, 1/3
for the $N2$ mapping of {\bf Figures 1 and 2}. Like Hoover and Posch, he assumed that Kaplan and Yorke's conjecture
for the information dimension was true. For the nonequilibrium $w$ values of 3, 4, and 5 he quotes information
dimensions 1.734, 1.506, and 1.376, as well as a general formula for the generalized Baker Maps. A decade later,
with Florian Grond\cite{b13}, we checked this assumption
for a flow, as opposed to a map.  We chose a four-dimensional chaotic
flow,
$$\{ \ \ddot q = -q - \zeta \dot q - \xi \dot q^3 \ ; \ \dot \zeta = \dot q^2 - T \ ; \ \dot \xi = \dot q^4 -3\dot q^2T \ \} ,
\longrightarrow D_{KY} = 2.80>D_I=2.56 \ .
$$
and soon discovered that the conjecture fails in that case.  For that four-dimensional chaotic problem, with a
relatively strong temperature gradient, $T = 1 + \tanh(q)$,
the interpolated Lyapunov sum, between those for two and for three exponents,
$\lambda_1+\lambda_2+ 0.80\lambda_3$ , vanishes. The consequent Kaplan-Yorke dimension, 2.80,
differs by about ten percent from the bin-based dimensionality of 2.56.
Those results, along with those that follow here leave the status of the conjecture perplexing. It
would be useful to have a clear informal  description of maps for which the conjecture is known to
be true accompanied by an illustrative list of situations where it fails.

\section{Area-Wise and Point-Wise Information Dimensions}

\begin{figure}
\includegraphics[width=3.6 in,angle=90.]{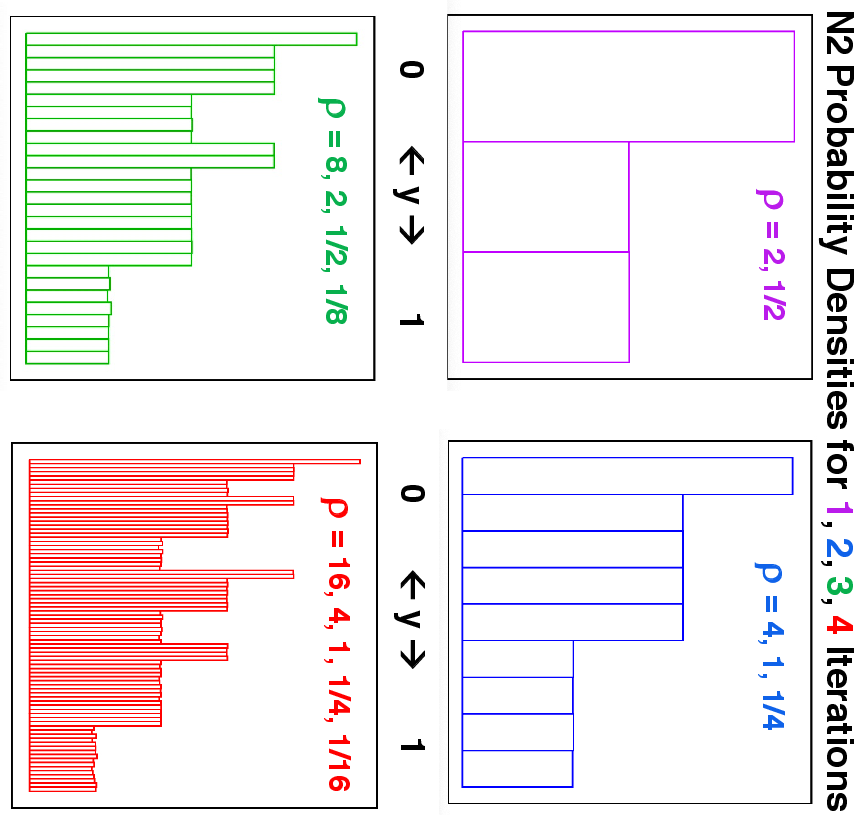}
\caption{
Histograms of the (base-4 logarithm of) probability density $\rho (y)$ for 1, 2, 3, and 4 area-wise iterations of
the $y$ component of the Baker Map $B$. Notice that the number of bins at each level of probability is the product
of a binomial coefficient and a power of two, in the red case $1\times 1$, $4\times 2$,  $6\times 4$,  $4\times 8$,
$1\times 16$. Notice here that the leftmost third of the interval, with summed-up probability 2/3, is reproduced as
a scale model (with the same information dimensionality) in the rightmost two thirds of the interval, with
probability 1/3. The information dimensions of all these iterates, $D_I =\sum (P \ln P)/\ln (\delta ) = 0.78969$ are
identical. This scale-model result differs from both the Kaplan-Yorke value of 0.7337 and the extrapolated 0.741
based on mesh sizes of $(1/3)^n$ and illustrated in {\bf Figure 4}. The histograms were constructed by binning (in
3, 9, 27, and 81 bins) the results of 1, 2, 3, and 4 iterations of 100,000 equally spaced initial values on the interval $0<y<1$. 
}
\end{figure}

Analyzing the fractal structures generated by the compressible $N2$ Baker Map reveals that there is {\it no fractal structure}
in the $x$ direction. See again the rotated maps' fractals in {\bf Figure 3}. Only the $y$ coordinate reveals a fractal.  This suggests two
computational approaches to determining the information dimension associated with the $y$ direction in map $B$ or
the $q=p$ direction in map $N2$: [ 1 ] Propagating a series of {\it area} mappings, starting
with a homogeneous square-lattice covering of the initial unit square or the rotated $2\times 2$ diamond-shaped
domain of {\bf Figure 1}; [ 2 ] Accumulating a time series of bin occupancies of {\it points}, with as many as
trillions of iterations generating a long sequence of {\it points} started at an arbitrary initial point.
Approaches [1] and [2], area-wise and point-wise, respectively appear to be equally legitimate routes to
information dimension. It was a surprise to find that the two don't agree although both these approaches do
reach well-defined limits. Another approach, [3], which we term ``stochastic'', adopts random numbers for
successive values of $x$ rather than using the more time-consuming analytic $N$ mapping. With random numbers
$\{ 0 < r < 1 \}$ the third approach is simply a confined random walk $(0 < y < 1)$ with red-region ``up'' steps
one-third of the time and ``down'' steps two-thirds of the time.
The programming of a single stochastic step requires two calls to a
random-number generator (for which we use a standard random-number FORTRAN subroutine). Note the underscore
in the ``calls'' below:
\begin{verbatim}
call random_number(r)
if(r.lt.1/3) y = (1+2y)/3
if(r.gt.1/3) y = (0+ y)/3
call random_number(x) for two-dimensional grid
\end{verbatim}
We have already seen, in {\bf Figure 4}, that the area-wise mapping used to generate the histograms,
simply repeats the single-iteration three-bin information dimension, 0.78969. The point-wise mapping
is simpler. It is only limited by available computer time. A personal computer is quite capable of
trillions of point-wise iterations. A billion point-wise iteration take about a minute of computer
time. Using double-precision and an initial point $(x,y) = (0.5,0.5) \leftrightarrow (q,p) = (0,0)$
the two algorithms, pointwise and stochastic agree, as expected, to four-figure accuracy, with the
following three-strip populations with a total of one billion points and the resulting entropies:
$$
{\rm pointwise}: 666 \ 681 \ 049 + 295 \ 151 \ 739 + 38 \ 167 \ 212 \rightarrow D = 0.6873 ;
$$
$$
{\rm stochastic}: 666 \ 631 \ 518 + 295 \ 178 \ 423 + 38 \ 190 \ 059 \rightarrow D = 0.6874 . 
$$
The close agreement shows that area-wise mapping is an outlier and suggests the adoption of
point-wise distributions. We consider some interesting details of that approach next.

\section{Point-Wise Information Dimension From The Baker Map Using a Random-Walk Algorithm}

\begin{figure}
\includegraphics[width=3.0in,angle=+90.]{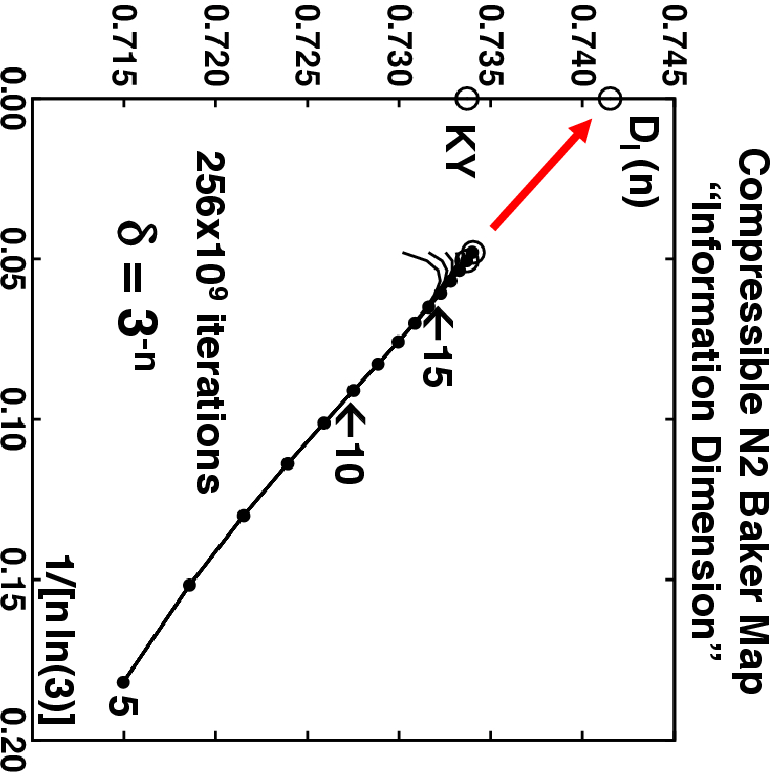}
\caption{
Stationary estimates of $D_I$ for the confined-random-walk model of the Baker Map with results for
$3^{5,10,15}$ equal bins emphasized. We saw above that the two values shown at the zero bin-size limit
($\delta \to 0$) are the Kaplan-Yorke dimension, 0.7337, and a plausible extrapolation of
trillion-iteration computations with as many as $3^{19}$ bins, $0.741_5$.  Note the qualitative
difference of the mesh dependence here ( the slope is uniformly negative here ) compared
to those shown for four other sets of bin sizes, $4^{-n}, \ 5^{-n}, \ 6^{-n}, \ 7^{-n}$ in
{\bf Figures 6 and 7}.  The two open circles at $n = 18$ and 19 correspond
to $1.024 \times 10^{12}$ iterations. The smaller dots correspond to sequences of 256 billion points.
}
\end{figure}

It is easy to verify that the one-dimensional and two-dimensional point-wise mappings agree with one
another for readily convergent simulations with $\delta = 3^{-10}$ or $3^{-15}$. Such results agree
very well with the stochastic map where ${\cal R}$ represents a random number from the interval $(0<x<1)$.
$$
{\cal R} < (1/3) \rightarrow y^\prime = (1+2y)/3 \ ; \ {\cal R} > (1/3) \rightarrow y^\prime = (y/3) \ .
$$
For a fixed choice of $\delta_y$ the three approaches agree to five-figure accuracy, supporting the use of
the simpler,  approach shown in {\bf Figure 5}. The data cover the range from $n= 5$ to $n=19$ with
the data approaching $D_I$ from below, eventually reaching a straight line with a well-defined limit $0.741_5$.

It is straightforward to write a supporting random-walk computer program distributing many
successive points over $3^n$ bins of width $(1/3)^n$. {\bf Figure 5} shows the results of distributing
up to a trillion iterations over as many as $3^{19} \simeq 10^9$ square bins. A single one-dimensional Baker-Map
mapping of a uniform distribution of ``many'' points ( millions or billions ) on the interval $(0 < y < 1)$
puts 2/3 of them into the lefthand interval of width $\delta = 1/3$. The remaining 1/3 of this singly-mapped
measure is mapped uniformly into the two remaining bins, center and right, of combined length 2/3.
{\bf Figure 4} illlustrates the iterated operation of the compressible
Baker Map for 1, 2, 3, and 4 iterations applied to an initially uniform distribution of 100000 points. For simplicity
here we have projected the result of the mapping onto the unit interval in $y$ rather than the
$2 \times 2$ diamond or unit square.  Propagating the singly-mapped measure results in measures of (2/3) and
(1/6) and (1/6) in the three equal-width bins, and so to an approximate single iteration information dimension,
after a single iteration of many uniformly-dense points gives
$$
D_I(1) = \langle {\rm P}\rangle/\ln (\delta)=
[ \ (2/3)\ln(2/3) + (1/6)\ln(1/6) + (1/6)\ln(1/6) \ ]/\ln(1/3) = 0.78969 \ .
$$
Here $\delta = 1/3$ is the bin size and the $\{$ P $\}$ are the probabilities of the three bins. The nine-bin
area-wise information dimension follows similarly with the leftmost bin probability of (4/9) followed by four
bins with probabilities (1/9) and four more with probabilities (1/36). Summing the nine P $\times$ ln(P) terms
and dividing by ln(1/9) gives {\it exactly the same} dimensionality as before, $D_I(2)= 0.78969$. Likewise
from the histogram data of {\bf Figure 4} for $D_I(3)=D_I(4)= 0.78969.$

Although initially it is
a surprise to find that the same information dimension results for 2 or 3 or 4 or \dots iterations, that result is
fully consistent with, and implied by, the scale-model nature of the distribution, as shown in {\bf Figure 4}. Iterating a
uniform coverage of the unit square or diamond suggests that the information dimension of the Baker maps
history is 1.78969. One would think that the limiting case $\delta \rightarrow 0$ would also result from
a long time series generated by point-wise iteration of a single point.  We saw in {\bf Figure 5} that
point-wise iteration suggests a different dimensionality, $1.741_5$!

\section{$N3$, A Well-Behaved Three-Panel Baker Map}

\begin{figure}
\includegraphics[width=3.0in,angle=+90.]{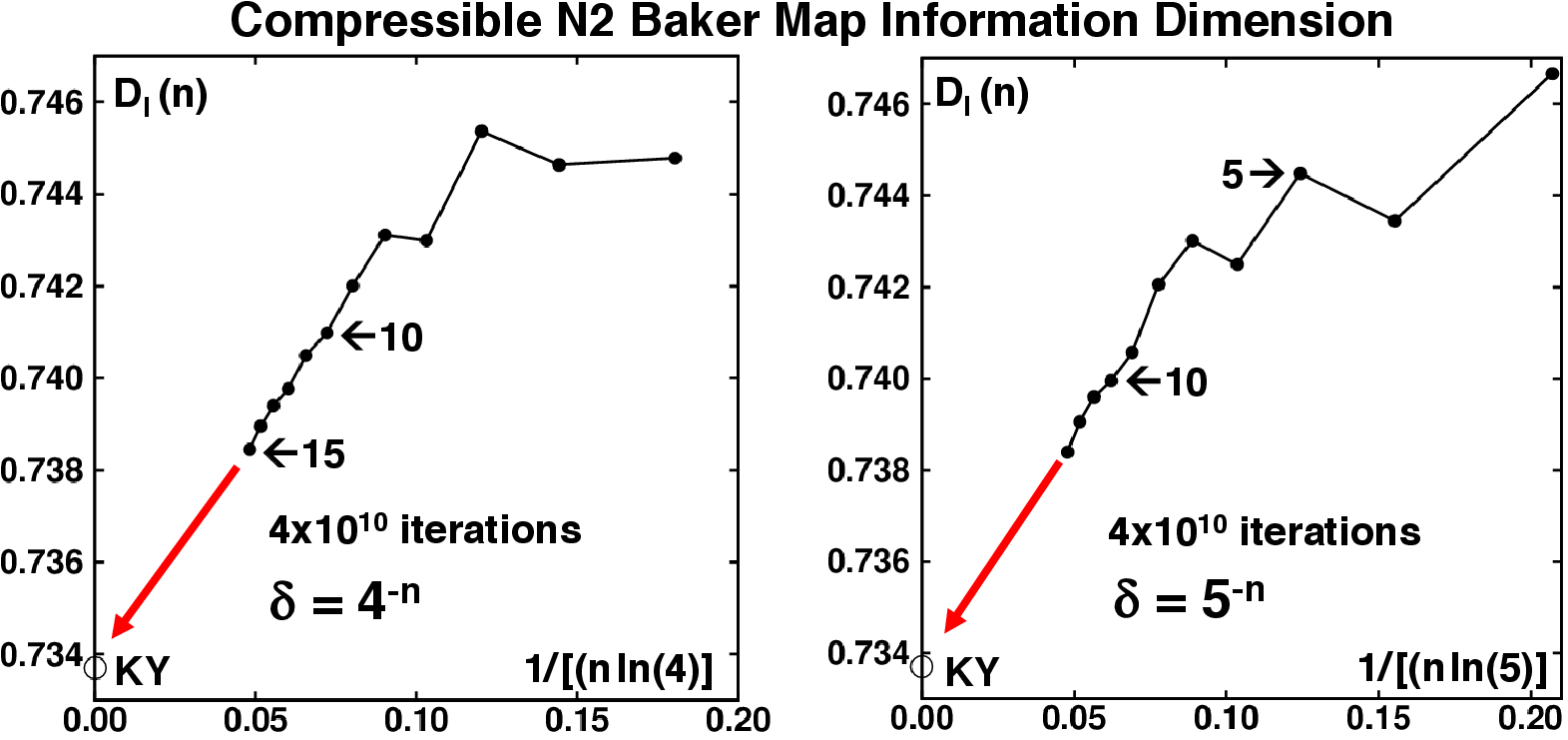}
\caption{
Stationary estimates for the Baker Map Information Dimension using up to $4^{15}$ and
$5^{13}$ bins of equal width. These data, based on forty billion iterations of the
random walk mapping suggest agreement with the Kaplan-Yorke dimensions of the
one-dimensional $y$ version of two-panel Baker Maps, $D_{KY} = 0.7337$.
}
\end{figure}

\begin{figure}
\includegraphics[width=2.0in,angle=+90.]{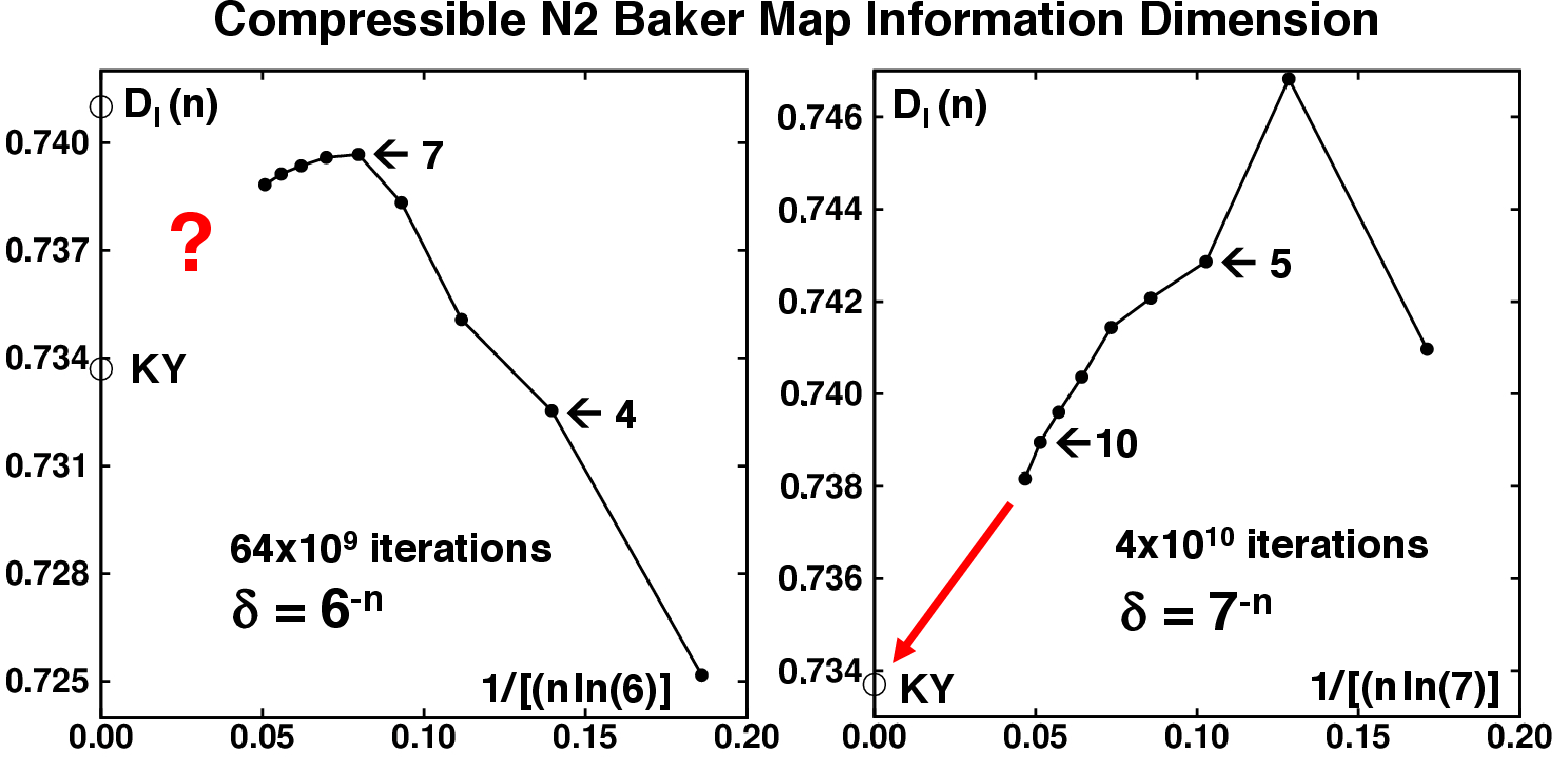}
\caption{
Stationary estimates for the one-dimensional version of the two-panel Baker Maps'
Information Dimensions using up to $6^{11}$ and $7^{11}$ bins of equal width. These data,
like those in {\bf Figure 4}, are based on forty billion iterations of the confined
random-walk mapping.  Both the Kaplan-Yorke dimension 0.7337 and the estimate 0.741 from
Reference 14, based on meshes with up to $3^{19}$ equal bins, are shown as open circles at
the left border of the $\delta = 6^{-n}$ plot.
}
\end{figure}

\begin{figure}
\includegraphics[width=2.3in,angle=+90.]{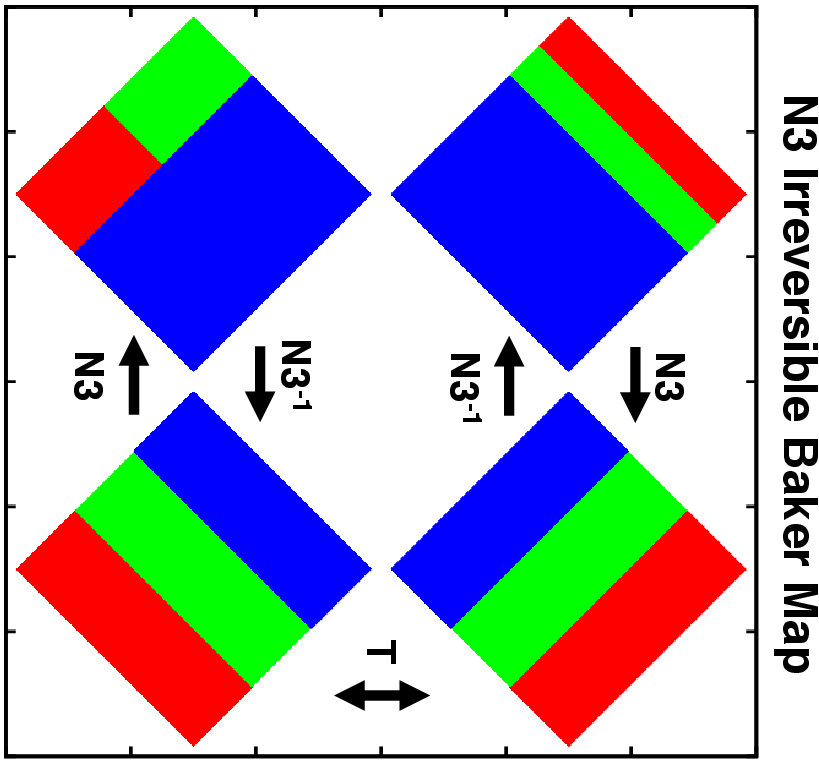}
\caption{
The three-panel Baker map $N3$ is slightly more complex than the two-panel $N2$ map, dividing
the upper left red and green portion in half. Applying the sequence of three maps $N3TN3$ shown
at the bottom left, is quite different to a mirror image of the original upper left.
Evidently the $N3$ map is not time-reversible. But both the area-wise and the point-wise
maps match the Kaplan-Yorke information dimension. Quite unlike the simpler two-panel Baker Map
$N2$ the three routes to the $N3$ information dimension all reach the same value $D_I = 0.78969$.
}
\end{figure}

Inspection of {\bf Figure 9} shows that with mapping both the red and green panels increase in width
by a factor 6 and decrease in length by a factor 3, while the blue panel, with probability 2/3,
increases by a factor 3/2 and decreases by a factor 3, giving rise to the Kaplan-Yorke dimension
$$
\lambda_1 = +0.867563 \ ; \ \lambda_2 = -1.09861231 \  \rightarrow
$$
$$
\lambda_1 + 0.78969\lambda_2=0 \rightarrow D_{KY} = +1.789690 \ ,
$$
the same as the information dimensions found with area-wise and point-wise analyses.
The probabilities associated with the $N3$ shown in the histograms of {\bf Figure 9} are identical to
those of $N2$, but with a different ordering of the histogram rectangles.
Evidently the $N3$ area-wise dimensionality, like the $N2$, doesn't change. Unlike $N2$ the $N3$ map does
agree with Kaplan-Yorke.
In a memoir for Francis Ree, Bill chose meshes from $(1/3)^5$ to $(1/3)^{18}$ for a set of $10^{11}$
iterations of the $N3$ map. His {\bf Figure 7} appears to be fully consistent with a point-wise estimation
$D_I = 0.79$.  Within the estimated uncertainty of 0.001 it appears that the $N3$ area-wise, point-wise,
and Kaplan-Yorke values of the information dimension all agree with one another. This makes the failure of
the simpler $N2$ Baker map, with only two linear panels, to provide simplicity a puzzling challenge.

Like $N2$ the three-panel $N3$ fractal corresponding to {\bf Figure 8} can be reproduced with calls to a random number generator. The
simplest program results if the $N3$ fractal is generated in the $y$ direction or in the two-dimensional
unit square, $0 < x,y < 1$:
\begin{verbatim}
call random_number(x)
             ynew = (1+y)/3 ! green
if(x.lt.1/6) ynew = (2+y)/3 ! red
if(x.gt.1/3) ynew = (0+y)/3 ! blue
call random_number(x)       ! if both (x,y) are desired
\end{verbatim}

\section{Conclusions and Discussion}

Relatively simple numerical work, on the order of a few dozen lines of FORTRAN, along with
a few hours of laptop time, are enough to characterize the variety of results for $D_I$
based on [1] iterating areas or [2] generating representative sequences of points. These
two different views of fractal structure are analogs of the Liouville and trajectory
descriptions of particle mechanics. We think the singular anisotropy of fractals favors the
pointwise approach. We found that pointwise analysis with the mesh series $(1/3)^n$ appears
to contradict the Kaplan-Yorke dimension while the alternative series $(1/4)^n,(1/5)^n,(1/7)^n$
appear to support it. The series $(1/6)^n$ is inconclusive.

Though the one-dimensional confined random walk provides a fractal distribution in $\{ y \}$,
indistinguishable from that for the compressible $N2$ Baker Map, the confined-walk analog lacks the
Baker-Map Lyapunov exponents on which the Kaplan-Yorke dimension relies :
$$
\lambda_1 = (1/3)\ln(27/4) \ ; \ \lambda_2 = (1/3)\ln(2/27) \to D_{KY} = 0.73368 \ .
$$
The variety of results obtained here for specific maps underlines the value of studying
particular, as opposed to general, models. There are several publications suggesting that
the information dimension is particularly robust to changes of variables\cite{b11}, certainly a
desirable property. On the other hand these results typically exclude mappings in which
infinitely many points where mapping discontinuities occur, a characteristic of Baker maps. 

Returning to the longstanding motivations of Loschmidt's Reversibility Paradox and Zerm\'elo's
Recurrence Paradox, compressible maps simplify our understanding of their resolutions, for flows just as well
as for maps. Fractal states have zero volume in their embedding spaces. Chaos provides exponentially
unstable (and therefore unobservable) repellors and exponentially stable (and therefore
inevitable) attractors. Time-reversible maps provide simple fractal examples of Second Law
irreversibility despite the paradoxes. Also notable is the quantitative agreement, within
Central Limit Theorem fluctuations, of reversible distributions with those generated using
stochastic random walks. Let us summarize the facts that stand out from our work: The simple
$N2$ two-panel map, whether one-dimensional, in $y$, or two-dimensional, in $(q,p)$, provides
three  different values of information dimension ``area-wise'', 0.78969 or 1.78969, ``point-wise'',
$0.741_5$ or $1.741_5$, and Kaplan-Yorke, 0.73368 or 1.73368. The more complex, but still linear,
three-panel $N3$ map is consistent with $D_I = 0.78969$ in one dimension and 1.78969 in two for
all three approaches.
\begin{figure}
\includegraphics[width=3.6 in,angle=+90.]{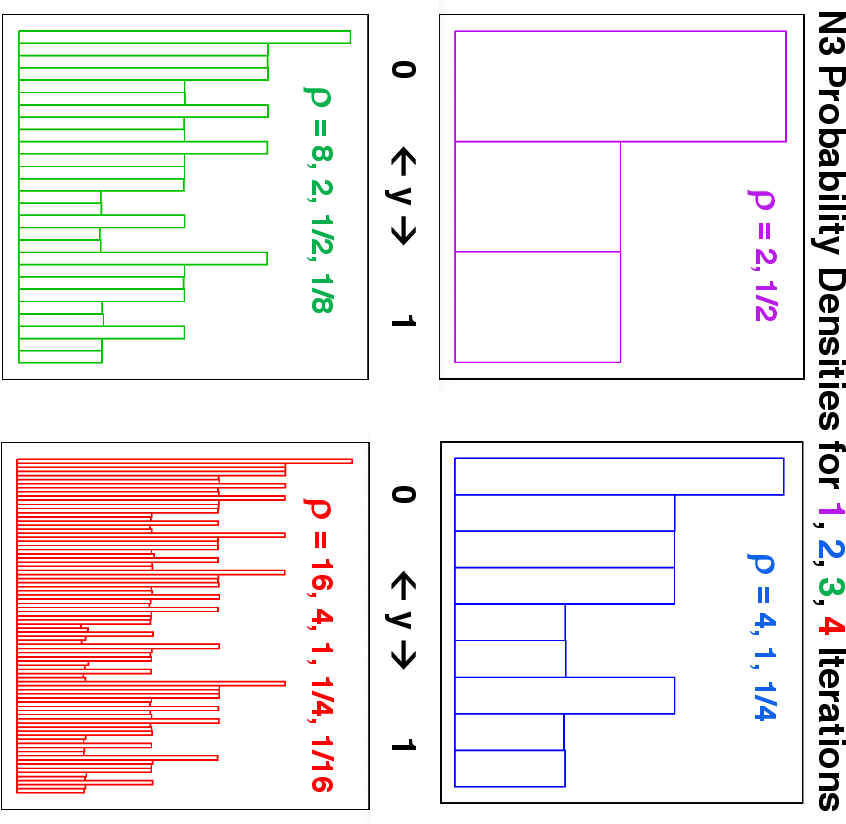}
\caption{
Histograms of the (base-4 logarithm of) probability density $\rho (y)$ for 1, 2, 3, and
 4 area-wise iterations of the $y$ component of the nonequilibrium $N3$ Map.  Notice that
the central and rightmost thirds of each resulting mapping are both perfect scale models
(reduced by a factor of four) of the leftmost third. This observation explains the
persistence of the three-bin information dimension throughout any number of iterations.
}
\end{figure}
\begin{figure}
\includegraphics[width=5.0 in,angle=+90.]{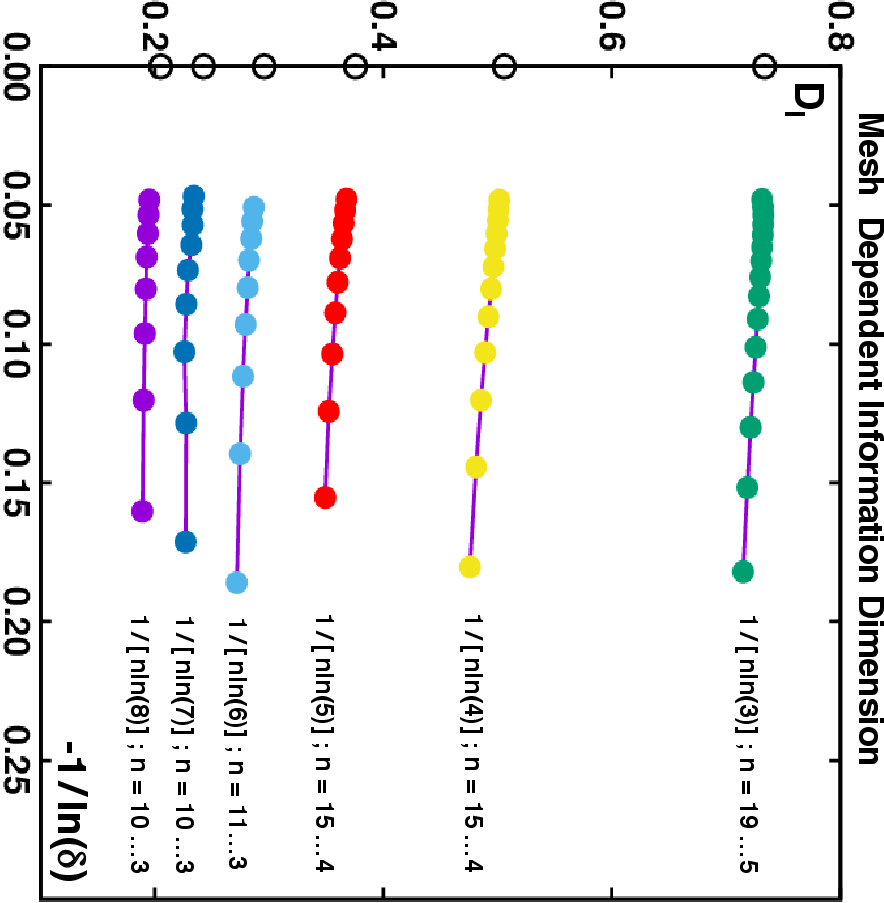}
\caption{
Dependence of the information dimension $D_I$ on the bin size $\delta$ for generalized
Baker Maps. The integers from 3 to 8 indicate the fraction of the unit square occupied by
the narrower rectangle of the mapping. As an example the $N2$ mapping of Figure 1 corresponds
to the integer 3. The points for each mapping give information dimensions for different
choices of the bin size, right-to-left from $(1/8)^3$ to $(1/8)^{10}$ for the bottom set
of eight data points and from $(1/3)^5$ to $(1/3)^{19}$ for the top set of green points.
The Kaplan-Yorke approximations are shown as open circles for each generalized map. They are
excellent approximations! Each filled circle is the result of two billion steps in a
random-walk simulation of the generalized Baker Map.
}
\end{figure}

\section{Acknowledgement}
Carl Dettmann,Thomas Gilbert, and Kris Wojciechowski kindly provided helpful advice and
references. Evidently the simplest generalized Baker Maps exhibit different area-wise and
point-wise information dimensions. Surprisingly, it took us 20 years to come to this
understanding. Thanks to our colleagues for their help along the way.

\section{Response to Comments by the Reviewers of March 2023}

 In early 2023  Tim Li requested that we contribute an article for a special issue of {\it Entropy} on
 ``Maximum Entropy Random Walks". This reminded us of our unpublished article of ours from
September 2019, "Random Walk Equivalence to the Compressible Baker Map and the Kaplan-Yorke Approximation
to Its Information Dimension". We brought that work up to date and submitted it to {\it Entropy}. Two
reviewer comments soon arrived, one enthusiastic and the other not. The unfavorable review suggested that
the manuscript had little to do with random walks or maximum entropy. We remark that our work was
described by the favorable reviewer as including a
novel, short, and highly-efficient random-walk algorithm to the evaluation of the "information dimension"
of maps chosen to shed light on the reversibility paradoxes of Loschmidt and Zerm\'elo. Kaplan and Yorke
formulated an approximate evaluation of the information dimension\cite{b12} from the entropy of a set of
points generated by an iterative solution of motion equations in the form of bin probabilities where the
bins span the space occupied by solution points:
$$
D_I = \sum P\ln P/\delta \simeq 1 - (\lambda_1/\lambda_2) = D_{KY} ,
$$ 
where $P$ is the occupation probability of a bin of width $\delta$ and $D_I$ is the information dimension,
a measure of the entropy of the set of points. Kaplan and Yorke's formula for $D$ is much discussed in the
literature, though the reasoning supporting it is obscure. We quote from page 169 of Tam\'as T\'el's and
M\'arton Grioz' excellent book, {\it Chaotic Dynamics}\cite{b18}:
\begin{quote}
Both the information dimension and the average Lyapunov exponents are determined by the natural
distribution. We can therefore expect to find an explicit relation between them.  This rule, called the
Kaplan-Yorke relation, is valid ... [ ! ] for chaotic attractors of general two-dimensional invertible map, and can
be obtained from a simple argument. [This is followed by two pages of informal text ending up with the
``valid" rule $D = 1 - (\lambda_1/\lambda_2)$.] \end{quote}
 Because this ``rule'' is violated by the $N2$ map we do not reproduce the text supporting it. As one
reviewer requests definitions of $P$ and the $\{ \ \lambda_i \ \}$ we reiterate that $P$ is the probability of occupying
a bin so that the sum over bins, $\sum P \equiv 1$, along with the usual formula for entropy,
$S \propto-\sum P\ln P$ provides the information dimension of the point set in the limit that the bin size
$\delta$ is small. The Lyapunov exponents, $N$ of them in an $N$-dimensional space, measure the exponential
growth and shrinkage rates of a small comoving ball in phase space, ordered from the largest, in average
value, to the most negative, so that a two-dimensional strange attractor set has a largest (positive)
Lyapunov exponent $\lambda_1$ as well as a negative exponent, $\lambda_2$, with the sum negative, signaling
the collapse of the set's dimensionality below 2. The Kaplan-Yorke formula provides a fractional
dimensionality between 1 and 2 (for a map) or 0 and 1 (for a confined random walk). In {\bf Figure 10} we 
display random-walk dimensions for six generalized Baker Maps (with mapping rectangles ranging from 1/3,
as in our Figures 1 and 2, and 1/8, where the latter mapping provides a sevenfold change in area. A discussion by
Doyne Farmer of similar map types can be found in Reference 20. Farmer analyzes the information dimension
of a map similar to our $N2$ map (in his Figures 2 and 4) but does not distinguish point-wise and area-wise
mappings as he was evidently unaware that the two can differ. For additional discussion of Lyapunov exponents and
information dimension we refer the reader to Reference 18 and to the corresponding Wikipedia articles on the web.

\end{document}